\documentclass[journal]{IEEEtran}
\usepackage{amsmath}
\usepackage{graphicx}

\ifCLASSINFOpdf

\else

\fi

\begin{document}
\bibliographystyle{IEEEtran}

\title{S$^{2}$MS:Self-Supervised Learning Driven Multi-Spectral CT Image Enhancement}
\author{Chaoyang~Zhang,
        Shaojie~Chang,
        Ti~Bai,
        and~Xi~Chen
\thanks{C. Zhang and X. Chen are with the School of Information and Communication Engineering, 
Xi’an Jiaotong University, Xi’an, Shanxi 710049, China (e-mail: xi\_chen@mail.xjtu.edu.cn)}
\thanks{S. Chang is with the Department of Radiology, Stony Brook University, Stony Brook, NY 11794, USA (e-mail: shaojiechang01@gmail.com)}
\thanks{T. Bai is with the Department of Radiation Oncology, University of Texas Southwestern Medical Centre, Dallas, TX 75390, USA}}

\maketitle

\pagestyle{empty}
\thispagestyle{empty}

\begin{abstract}
Photon counting spectral CT (PCCT) can produce reconstructed attenuation maps in different energy channels, reflecting energy properties of the scanned object. Due to the limited photon numbers and the non-ideal detector response of each energy channel, the reconstructed images usually contain much noise. With the development of Deep Learning (DL) technique, different kinds of DL-based models have been proposed for noise reduction. However, most of the models require clean data set as the training labels, which are not always available in medical imaging field. Inspiring by the similarities of each channel's reconstructed image, we proposed a self-supervised learning based PCCT image enhancement framework via multi-spectral channels (S$^{2}$MS). In S$^{2}$MS framework, both the input and output labels are noisy images. Specifically, one single channel image was used as output while images of other single channels and channel-sum image were used as input to train the network, which can fully use the spectral data information without extra cost. The simulation results based on the AAPM Low-dose CT Challenge database showed that the proposed S$^{2}$MS model can suppress the noise and preserve details more effectively in comparison with the traditional DL models, which has potential to improve the image quality of PCCT in clinical applications.
\end{abstract}


\IEEEpeerreviewmaketitle

\section{Introduction}

\IEEEPARstart{P}{HOTON} counting spectral CT (PCCT) can separately collect the incident photons in different energy bins, which has high energy resolution and can generate more accurate material decomposition \cite{r1,r2}. Nevertheless, with the increase number of energy bins, counting rate is limited in each individual channel, which results in a relatively low signal-to-noise ratio (SNR). Moreover, there are complicated noises caused by non-ideal response of detector, such as fluorescence x-ray effects, K-escape, charging sharing, and pulse pileups \cite{r3}. Noise in the reconstructed CT images will seriously affect diagnosis of doctors.

To reduce noise in CT images, recent deep learning (DL) technique has been widely developed in the field of CT image denoising and shows the potential in applications. Yang et al.
used the generative adversarial network (GAN) with Wasserstein distance and perceptual similarity to reduce the noise in CT images \cite{r4}. Lv et al. proposed an PCCT image denoising method via fully convolutional pyramid residual network, which suppresses noise in each single energy channel image \cite{r5}.

However, traditional DL methods require high-quality clean images as training labels to achieve high performance, which are difficult to obtain especially in medical imaging field. To solve this problem, Lehtinen et al. introduced the Noise2Noise model (N2N), where the network was trained to map one noisy realization to another noisy realization \cite{r6}. In addition, photon counting spectral CT (PCCT) provides an opportunity to produce reconstructed attenuation maps in different energy channels, which reflect energy properties of the scanned object. Using the similarity of images in different energy bins, we proposed a self-supervised learning driven PCCT image denoising method via multi-spectral channels based on the N2N network model (S$^{2}$MS). In our framework, the input are images from multi-channels and a channel-sum image, while the output is image of one single channel. Both input and output are noisy PCCT images. Compared with N2N, our proposed S$^{2}$MS fully used all the reconstructed images in different energy bins at the same time, rather than processing images in each channel separately. All the simulated experiments were carried out and the results show the S2MC model is effective and accurate in noise reduction and detail preservation.

\section{Materials and Methods}
\subsection{Basic Principle of PCCT}
Compared with the traditional energy integration detector, photon counting detector (PCD) can separately count out the number of photons in each energy channel, which can be used to obtain the projection in different energy bins. Fig. 1 shows an example of PCCT images in four energy channels. The similarity between images in different energy bins can be used for noise reduction.

\begin{figure}[ht]
\centering
\includegraphics[scale=0.18]{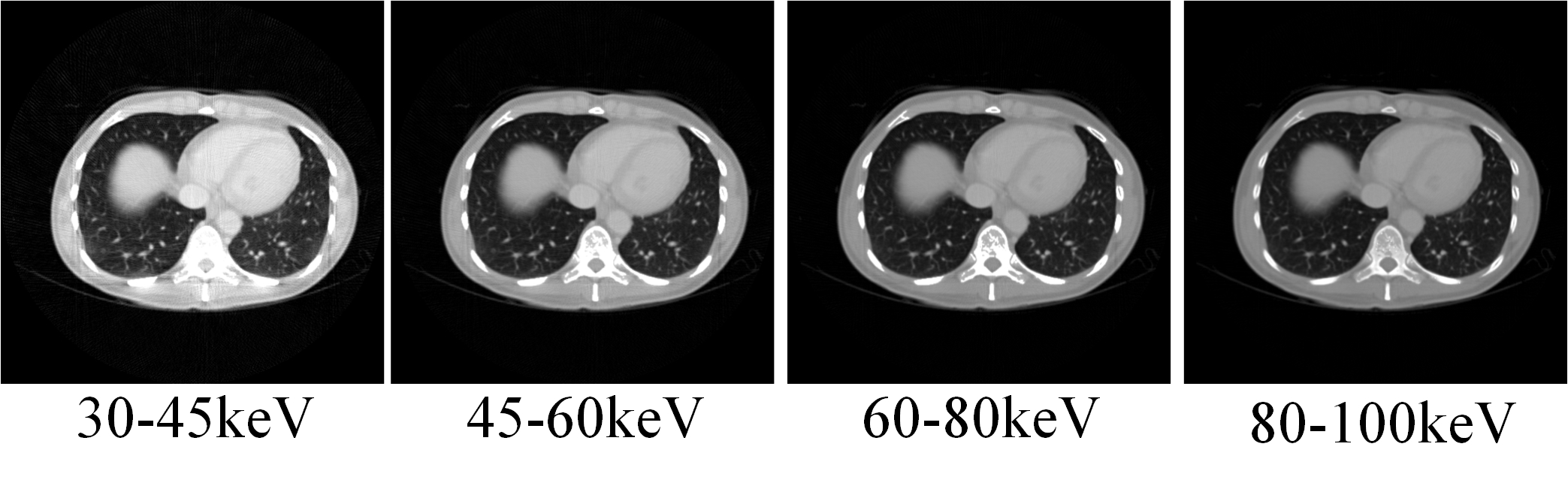}
\caption{An example of reconstructed PCCT images in different energy bins. The display window is [0,0.4] cm$^{-1}$.}
\end{figure}

\begin{figure*}[ht]
\centering
\includegraphics[scale=0.3]{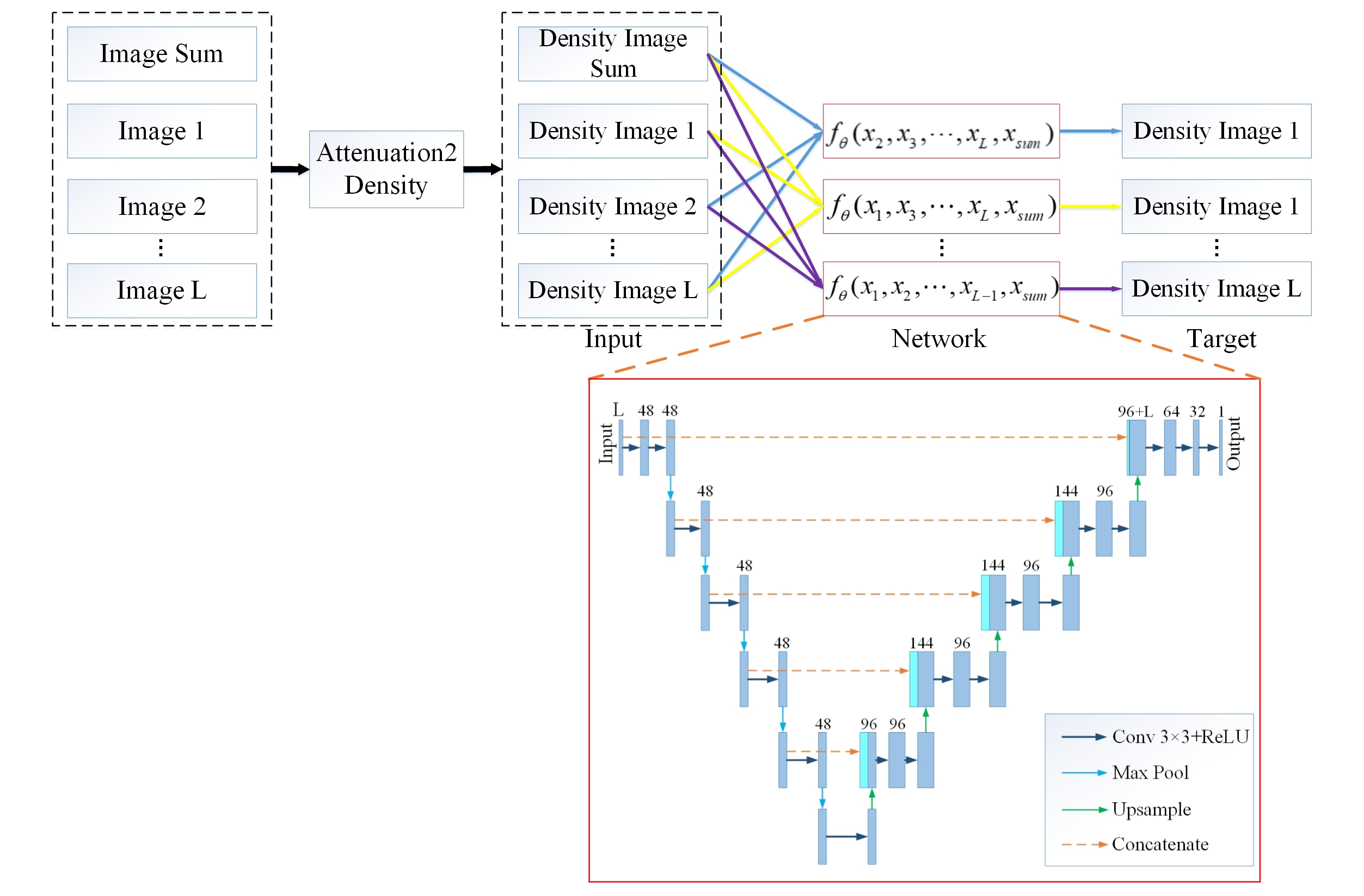}
\caption{The noise2noise network-based PCCT image denoising framework with self-supervised learning via multi-spectral channels (S$^{2}$MS). The attenuation images were divided by the mass attenuation coefficient, converted into density images.}
\end{figure*}

\subsection{Deep-Learning based Denoising}
In deep learning based denoising method, the input of the network is generally regarded as the following:
\begin{equation}
x_i=y_i +n_i
\end{equation}
where $x_i$ is the corrupted input, $y_i$ is the clean target, and $n_i$ denotes the corresponding noise. According to the type of training target, DL-based denoising can be divided into the following two types:
\subsubsection{Supervised Learning-N2C}

Traditionally, supervised learning is always used for deep-learning based CT image denosing, which means training a regression model with pairs $(\hat{x_i},y_i)$ and minimizing the function:
\begin{equation}
\theta^{*}=\mathop{\arg\min}\limits_{\theta}\frac{1}{N}\sum\limits_{i}\|f_{\theta}(\hat{x_i})-y_i\|
\end{equation}
where $f_{\theta}(\hat{x_i})$ is the denoising convolutional neural network (CNN), $\theta$ is weight, $N$ is the total number of training samples. In this paper, the image denoising based on supervised learning is referred as Noise2Clean (N2C).

\subsubsection{Self-Supervised Learning-N2N}

Opposite to Noise2Clean, Noise2Noise (N2N) is a self-supervised learning framework where input and target are both corrupted. It can be expressed as:
\begin{equation}
\theta^{*}=\mathop{\arg\min}\limits_{\theta}\frac{1}{N}\sum\limits_{i}\|f_{\theta}(y_i+n_{i1})-(y_i+n_{i2})\|
\end{equation}
where $n_{i1}$ and $n_{i2}$ are two independently noise realizations. It has been assumed that the Noise2Noise training is equivalent to Noise2Clean training under certain mild conditions \cite{r6,r7}:

$1. N\rightarrow\infty;$

$2. Conditional~expection~E\{n_{i2}|y_i\}=0;$

$3. n_{i1}~and~n_{i2}~are~independent;$

$4. {\forall}i, |f_\theta(y_i+n_i)| \l\infty.$

Since filter in the convolutional neural network is shift-invariant, different parts of the image may be served as multiple training samples. Even if the size of training data is small, the actual number of training samples is large enough to satisfy condition 1. After reconstructing, the noise in image domain is zero-mean and independent in different energy channels \cite{r8}, which means condition 2 and condition 3 are both satisfied in our method. Condition 4 can be easily satisfied by choosing the meaningful parameters of the network.

\subsection{Noise2Noise Network for PCCT Image Denoising}
By using the similarities of reconstructed images in different energy bins, we proposed a noise2noise network-based PCCT image denoising framework based on self-supervised learning via multi-spectral channels (S$^{2}$MS). The basic process of our framework is shown in Fig. 2.

In S$^{2}$MS, L-1 reconstructed images in single channel and a channel-sum image (linear attenuation map) were divided by the mass attenuation coefficient of each channel (Attenuation 2Density), converted into density images before training, which were used as the input of L channels. Then, the left single channel image was also converted into density image as the target. The S$^{2}$MS network can be described as:
\begin{equation}
\begin{aligned}
\theta^{*}=\mathop{\arg\min}\limits_{\theta}\frac{1}{N}\sum\limits_{i}\|f_{\theta}(y_{i1}+n_{i1},\cdots,y_{i(E-1)}\\+n_{i(E-1)},y_{isum}+n_{isum})-(y_{iE}+n_{iE})\|
\end{aligned}
\end{equation}
where $f_{\theta}(\hat{x_i})$ is the denoising convolutional neural network (CNN) with L inputs, $y_{isum}$ is the clean channel-sum image, $n_{isum}$ is the noise in channel-sum image,
$y_i={y_{i1},y_{i2},\cdots,y_{iL}}$ are the clean reconstructed images in different single energy channel and $n_i={n_{i1},n_{i2},\cdots,n_{iL}}$ are the corresponding noise.

After training, S$^{2}$MS can denoise the PCCT image in single channel. The denoised density image was multiplied by mass attenuation coefficient (Density2Attenuation), converted into a linear attenuation image (Fig. 3).

\begin{figure}[ht]
\centering
\includegraphics[scale=0.25]{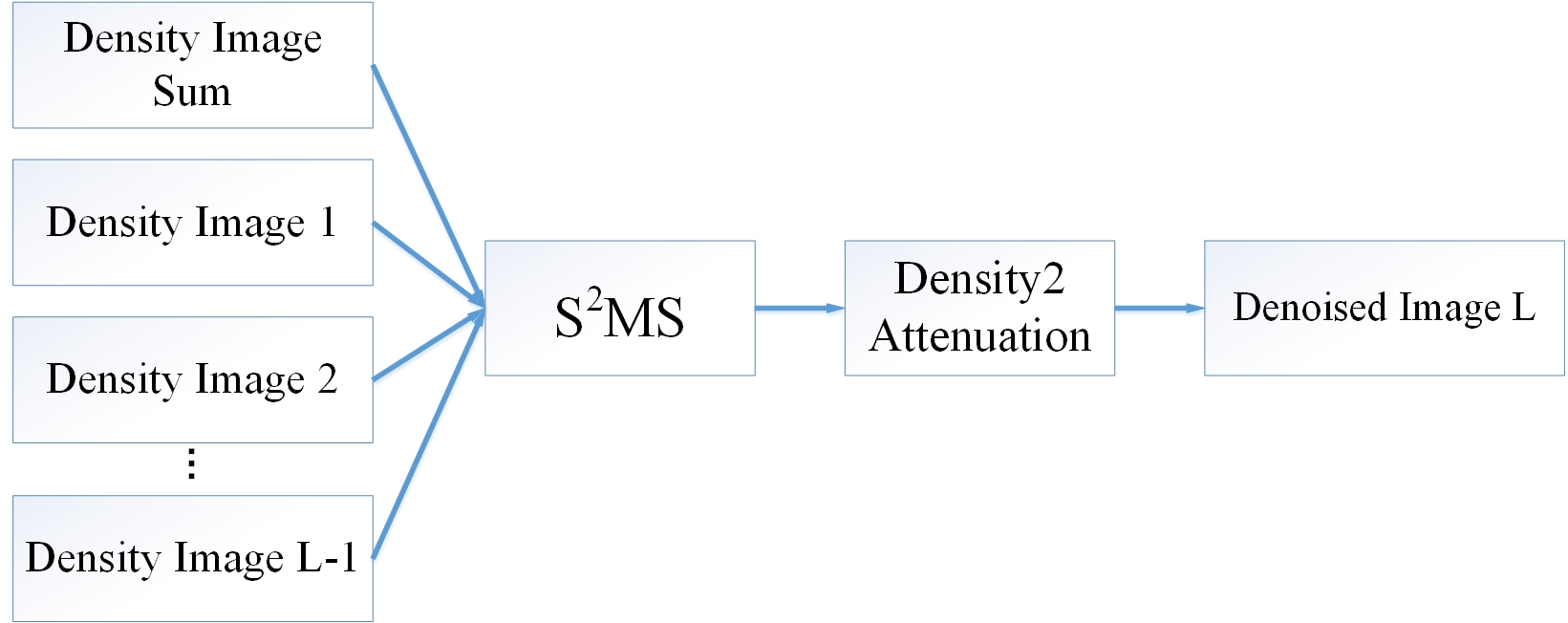}
\caption{The denoised process of the trained S$^{2}$MS network.}
\end{figure}

\subsection{Experiment Setup}
\subsubsection{Dataset Establishment}
In this study, CT images from the 2016 Low-dose CT Grand Challenge dataset \cite{r9} were used to simulate the PCCT images. 1000 slices of 7 patients were randomly divided into training dataset, validation dataset and test dataset according to the ratio of 8:1:1. An equal spatial fan-beam geometry was assumed to simulate the projecting process. The distance from the source to the system origin was 142 cm, the distance from the source to the detector was 180 cm, and there were 512 detector elements with the width of 0.1 cm per element. A total of 512 projections were acquired in an angular range of 360 degrees. The projection data were collected in four different energy bins 30-45keV (channel 1), 45-60keV (channel 2), 60-80keV (channel 3), and 80-100keV (channel 4). In each energy channel, 1000 PCCT images were acquired which had 512*512 pixels. Poisson noise was introduced in the simulation process. Totally   photons emitted along each x-ray path and the number of photons per energy channel was proportional to the normalized spectrum of each channel. Finally, the PCCT images can be reconstructed by FBP algorithm. Before training, the reconstructed images divided by the mass attenuation coefficient to convert into the density images which were used as the input and target of S$^{2}$MS.
\subsubsection{Network Implementation}
U-Net architecture in [6] was used in our study (Fig. 2). The encoder-decoder network includes a shrinking multi-scale decomposition path and a symmetric expansion path, with skip connections on each layer. Adam optimizer was used with a learning rate of 0.0003. The loss function was designed based on Mean Squared Error (MSE):
\begin{equation}
MSE(x,y)=\frac{1}{mn}\sum\limits_{i=1}^{m}\sum\limits_{j=1}^{n}\|x(i,j)-y(i,j)\|^2
\end{equation}
where $x$ is input and $y$ is target of the network.

The training was performed on a server with Intel Xeon Silver 4214 CPU and GeForce RTX 3090 24G GPU. The network was coded in Pytorch1.9.1 using Ubuntu20.04.
\subsubsection{Comparison Study}
To evaluate the performance of our proposed method, the Noise2Clean (N2C) and traditional Noise2Noise (N2N) were used for comparison. In N2C network, PCCT image reconstructed from the projections without noise was used as the target. In N2N network, two independent projections were simulated and the corresponding reconstructed images were used as input and output, respectively. Since our study focused on a denoising method rather than the network structure, the U-Net architecture in Fig. 2 was also used for N2C and N2N.
\subsubsection{Evaluation Metrics}
In our study, structure similarity (SSIM) and Root Mean Squared Error (RMSE) were used as the evaluation metrics. SSIM measures the structural similarity by comparing both the mean value and distribution relevance between denoised image and reference, and RMSE measures the L2-norm error between the estimated image and the ground truth.

\section{Results}
Denoised images generated by our proposed method S$^{2}$MS, N2C and N2C were shown in Fig. 4. Our proposed method can effectively reduce noise in each PCCT channel image. Especially in channel 3 and channel 4, our proposed method is able to retain richer structural information while suppressing most noise. Four regions of interest (ROIs) were selected (red rectangles) to show the detail preservation performance (Fig.5). In comparison with other methods, the proposed S$^{2}$MS can remove more noise while preserving more details.
\begin{center}
\begin{table}[h]
\centering
\caption{DENOISING RESULTS OF DIFFERENT METHODS ON TEST DATASET}
\setlength{\tabcolsep}{6mm}{
\begin{tabular}{llll}
\hline
Energy     & Method & SSIM   & RMSE   \\ \hline
           & N2N    & 0.9754 & 0.0131 \\
30-45 keV  & N2C    & 0.9765 & 0.0130 \\
           & S$^{2}$MS   & 0.9800 & 0.0117 \\ \hline
           & N2N    & 0.9738 & 0.0087 \\
45-60 keV  & N2C    & 0.9750 & 0.0085 \\
           & S$^{2}$MS   & 0.9853 & 0.0061 \\ \hline
           & N2N    & 0.9697 & 0.0072 \\
60-80 keV  & N2C    & 0.9719 & 0.0070 \\
           & S$^{2}$MS   & 0.9825 & 0.0048 \\ \hline
           & N2N    & 0.9442 & 0.0081 \\
80-100 keV & N2C    & 0.9479 & 0.0077 \\
           & S$^{2}$MS   & 0.9747 & 0.0046 \\ \hline
\end{tabular}}
\end{table}
\end{center}

We selected a fixed size (200*200) ROI (blue rectangle in Fig. 4) and calculated the SSIM and RMSE (Table I). Our proposed S$^{2}$MS achieved the highest SSIM and the lowest RMSE in each energy channel, which indicated that S$^{2}$MS made a better performance on image denoising for PCCT in comparison with N2N and N2C.

\begin{figure}[ht]
\centering
\includegraphics[scale=0.15]{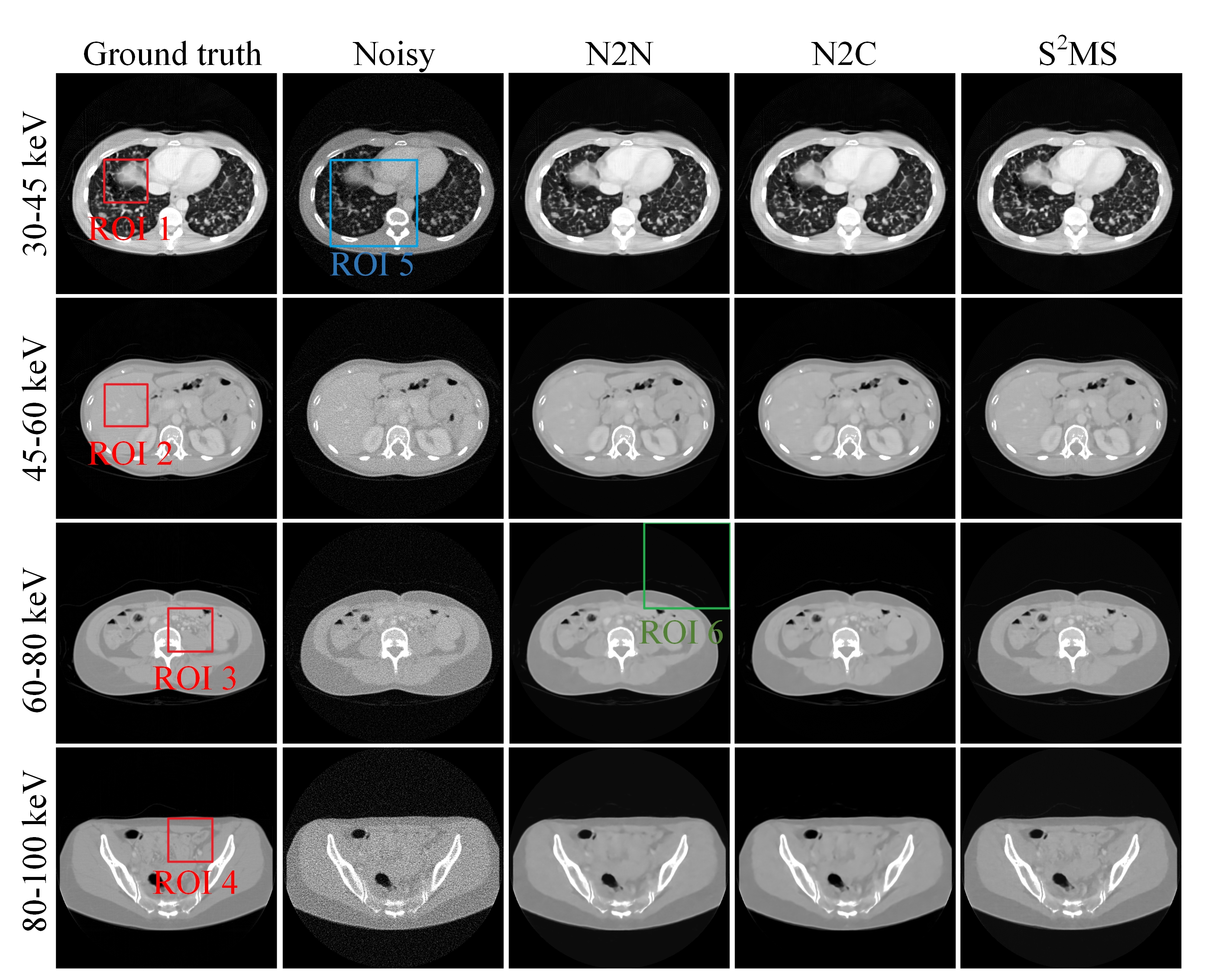}
\caption{Four example reconstructed slices in four energy channels (30-45keV, 45-60keV, 60-80keV, 80-100keV). The display windows for linear attenuation from the top to the bottom rows are [0,0.4] cm$^{-1}$, [0,0.4] cm$^{-1}$, [0,0.35] cm$^{-1}$, and [0,0.35] cm$^{-1}$, respectively.}
\end{figure}

\begin{figure}[ht]
\centering
\includegraphics[scale=0.15]{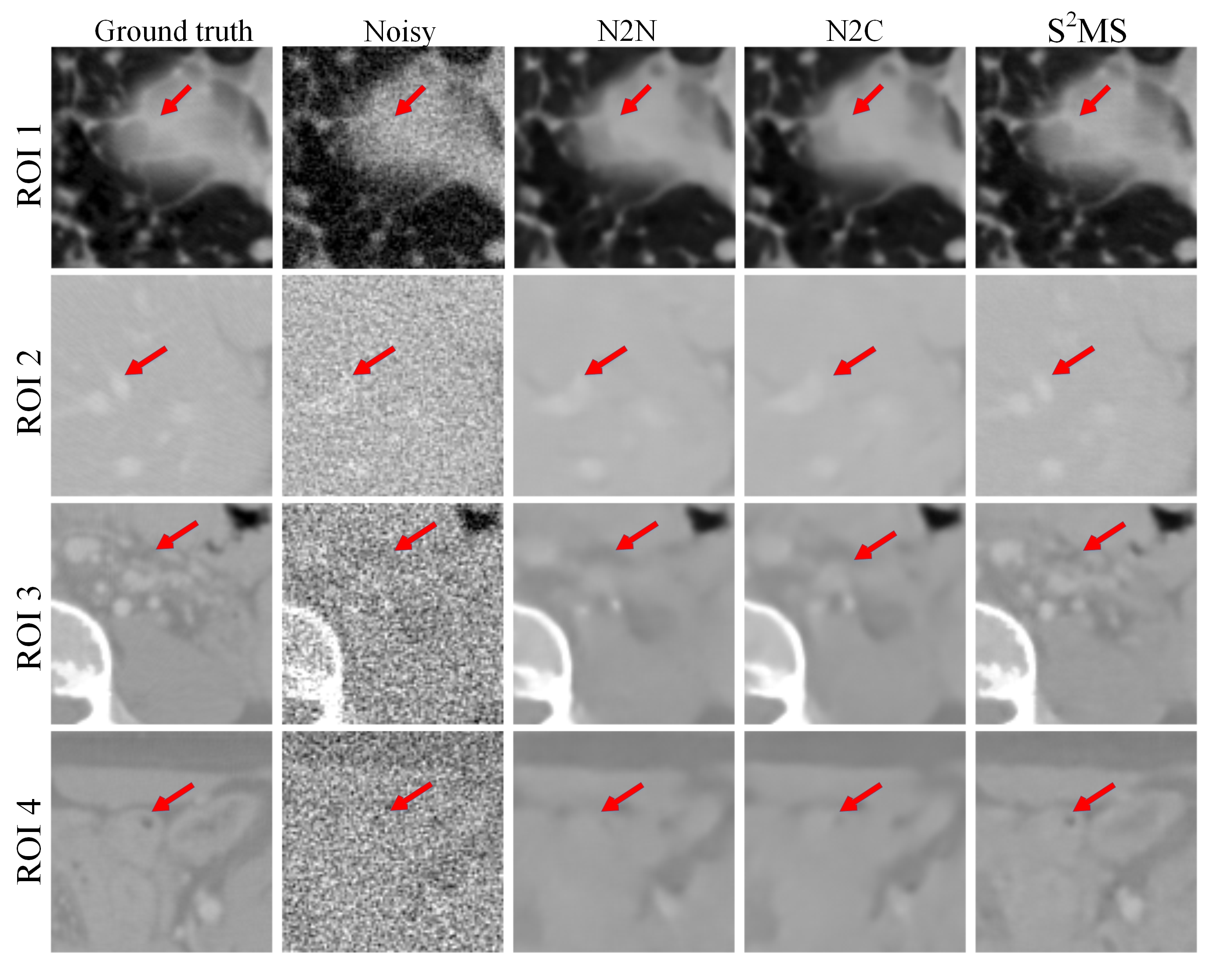}
\caption{Details of reconstructed images in Fig.3. The first and second rows are PCCT images in channel 3 (60-80 keV), the third and fourth rows are PCCT images in channel 4 (80-100 keV), and the display window for all images is [0,0.35] cm$^{-1}$.}
\end{figure}

\begin{figure}[ht]
\centering
\includegraphics[scale=0.15]{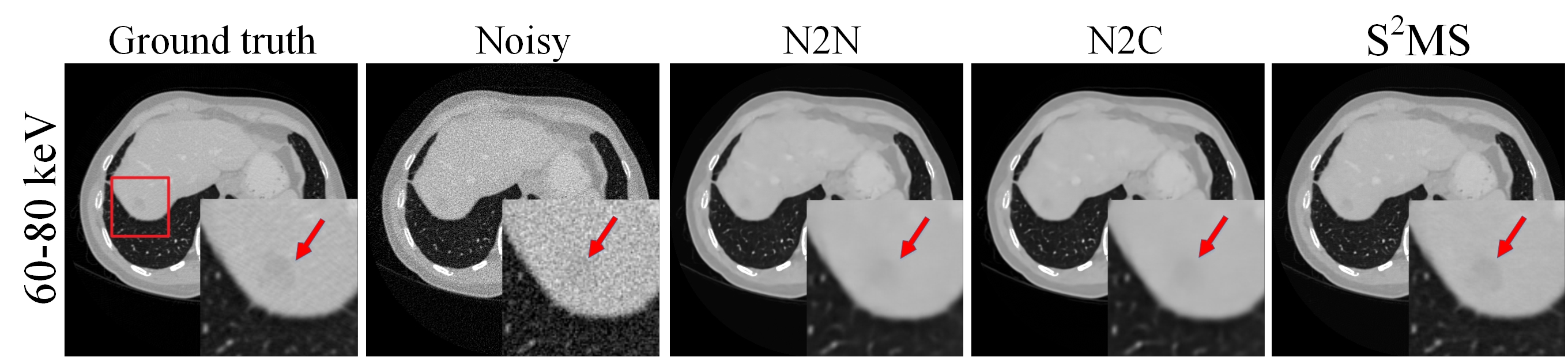}
\caption{A reconstructed PCCT image in channel 3, the noisy image and the outputs of different network. The shading in the red rectangle is the region of lesion which is magnified and shown. The display window is [0,0.35] cm$^{-1}$.}
\end{figure}

\begin{figure}[ht]
\centering
\includegraphics[scale=0.23]{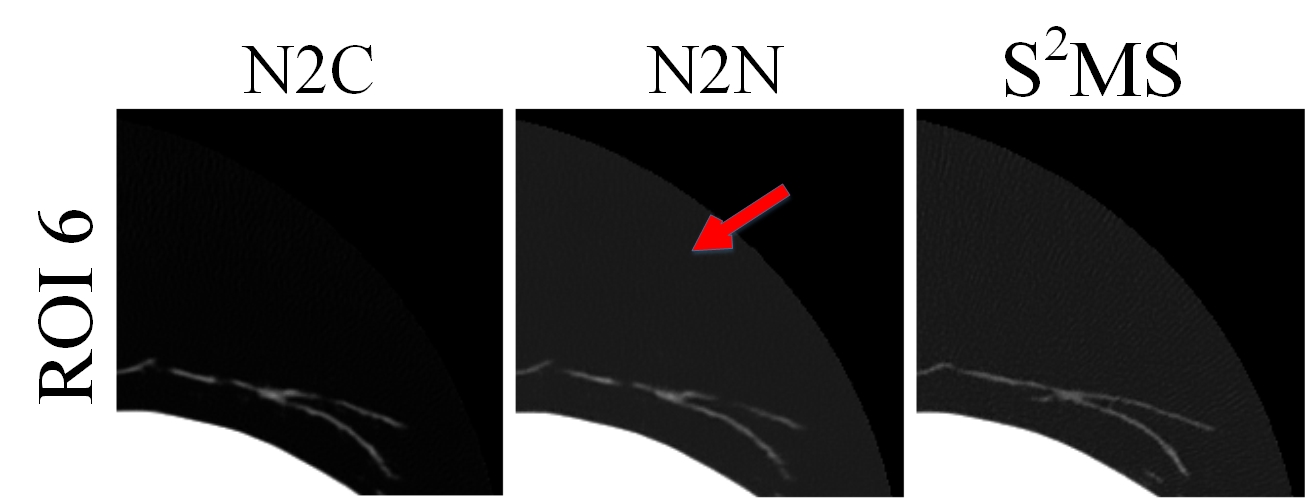}
\caption{The magnified image of ROI (green rectangle in Fig. 3). The display window is [0,0.1] cm$^{-1}$.}
\end{figure}

To further validate the performance of the proposed method in PCCT clinical application, CT images with lesion is shown in Fig. 6. Denoised reconstructed images in channel 3 (60-80 keV) of different methods are illustrated and the lesion region (red arrow) is magnified. The lesion is hardly observed in the noisy image while it can be clearly in the reconstructed image by S$^{2}$MS. The lesion area was blurred in the reconstructed images denoised by other methods. This result indicated that our method has potential in clinical application.

\section{Discussion and Conclusion}
In particular, in the output of N2N and S$^{2}$MS, a gray circular shadow can hardly be seen in the air area (Fig. 7). There is no anatomic structure in this area, which makes the noise distribution and background signal in this part are totally different from those in human part. Therefore, the output of the network may get wrong values in non-human regions. The shadow is hard to see and has little influence in diagnosis.

In conclusion, we have proposed a Noise2Noise-based PCCT image denoising framework via multi-spectral channels (S$^{2}$MS). In this study, noisy PCCT images were used as both the input and the output to train the network. To make full use of the spectral data in L channels, the reconstructed images in L-1 single channels and channel-sum image were used as the input and the left single channel image was used as the output. Compared with the traditional DL denoising method, simulation results show that the proposed method can obtain a reconstructed image with high quality: noise is reduced remarkably and detail features is well remained. No clean image needed makes the proposed S$^{2}$MS has potential in practical application. In the future work, our S$^{2}$MS will be regarded as a priori information to be combined with the material decomposition framework and the experimental data will be used to test the network.

\ifCLASSOPTIONcaptionsoff
  \newpage
\fi

\bibliography{refer}

\end{document}